\documentclass{article}

\usepackage{arxiv}

\usepackage[utf8]{inputenc} 
\usepackage[T1]{fontenc}    
\usepackage{hyperref}       
\usepackage{url}            
\usepackage{booktabs}       
\usepackage{amsfonts}       
\usepackage{nicefrac}       
\usepackage{microtype}      
\usepackage{lipsum}
\usepackage{graphicx}
\usepackage{subfigure}
\graphicspath{ {./images/} }
\usepackage{amsmath}
\usepackage{multicol}
\usepackage{multirow}

\title{EFFICIENT QUANTUM IMAGE REPRESENTATION AND COMPRESSION CIRCUIT USING ZERO-DISCARDED STATE PREPARATION APPROACH}

\author{
Md Ershadul Haque \\
  School of Computing, Mathematics and Engineering\\
  Charles Sturt University\\
  Bathurst, NSW 2795 \\
  \texttt{mhaque@csu.edu.au} \\
   \And
Manoranjan Paul \\
  School of Computing, Mathematics and Engineering\\
  Charles Sturt University\\
  Bathurst, NSW 2795 \\
  \texttt{mpaul@csu.edu.au} \\
  \And
 Anwaar Ulhaq \\
  School of Computing, Mathematics and Engineering\\
  Charles Sturt University\\
  Bathurst, NSW 2795 \\
  \texttt{aulhaq@csu.edu.au} \\
    \And
Tanmoy Debnath \\
  School of Computing, Mathematics and Engineering\\
  Charles Sturt University\\
  Bathurst, NSW 2795 \\
  \texttt{tdebnath@csu.edu.au} \\
}
\begin{document}
\maketitle
\begin{abstract}
Quantum image computing draws a lot of attention due to storing and processing image data faster than classical. With increasing the image size, the number of connections also increases, leading to the circuit complex. Therefore, efficient quantum image representation and compression issues are still challenging. The encoding of images for representation and compression in quantum systems is different from classical ones. In quantum, encoding of position is more concerned which is the major difference from the classical. In this paper, a novel zero-discarded state connection novel enhance quantum representation (ZSCNEQR) approach is introduced to reduce complexity further by discarding '0' in the location representation information. In the control operational gate, only input '1' contribute to its output thus, discarding zero makes the proposed ZSCNEQR circuit more efficient. The proposed ZSCNEQR approach significantly reduced the required bit for both representation and compression. The proposed method requires 11.76\% less qubits compared to the recent existing method. The results show that the proposed approach is highly effective for representing and compressing images compared to the two relevant existing methods in terms of rate-distortion performance.  
\end{abstract}
\section{Introduction}
\label{sec:intro}

In quantum computing, quantum mechanic properties such as superposition, entanglement, and parallelism play a vital role in faster computation than classical in terms of computation, storage, and complexity \cite{b1,b2,b3,b4,b5}. In the quantum image, qubits replace the classical bits in an array of pixels \cite{b10}. 
In \cite{b5,b6}, Shor and Grover proposed an approach for factorial calculation and database search which is faster than classical \cite{b6}. Recently, quantum image representation and compression got a lot of attention due to faster computation in terms of image processing are image representation, image compression, watermarking, information security, and high privacy \cite{b5,bb8,bb910}.\\
Most of the existing models connect the pixel values representing qubits to the position values representing qubits directly \cite{b8}. For this reason, a huge number of gates is necessary to complete the connection to present a real image. Nasr et al. proposed an EFRQI(efficient flexible representation of the quantum image) approach to address higher operational gates by adding an auxiliary qubit that makes a connection between pixel and state representing qubits. Although, it reduces bits compared to NEQR (novel enhanced quantum representation) approach, still, this approach requires a significant number of gates \cite{b9}. To address the higher bit rate issue of an EFRQI, haque et al. proposed a novel SCMFRQI (state connection modification FRQI) approach \cite{haque2022novel} that uses only one reset gate rather than twice uses of Toffoli gate to reduce the number of required gates \cite{b9}.\\
In this work, we propose a ZSCNEQR approach that drops the zero connection from the state label preparation via $8\times8$ block-based processing. After discarding '0', the proposed ZSCNEQR circuit work fine because '1' is the main element that flip the output state of the operational gate as the '0' is not the control element. Also, the quantum transformation of '0' can be designed using a quantum identity gate which can be ignored in the quantum circuits as NEQR \cite{bb10} did for the pixel value. This ignorance has no effect on the overall quantum circuit performance, however, this decreases the quantum circuit complexity. The proposed model is tested and verified using the benchmark simulator. The experiments are conducted to see the performance of the image representation. Moreover, the compression performance is also conducted by applying the proposed representation in the quantized DCT transformed data. The results show that the proposed ZSCNEQR approach performs superior to saving the required number of transmitted bits compared to SCMEFRQI and EFRQI approaches \cite{haque2022novel}. The rest of the article is organized as follows. The literature survey is described in Section \ref{L_R}; the proposed methodology is outlined in Section \ref{P_M}; the result and its description are given in Section \ref{R_D}. The conclusion of this work is summarized in section \ref{CC}.

\section{Literature survey}\label{L_R}
\begin{figure}[htbp]
\centerline{\includegraphics[width=\linewidth,height=10cm]{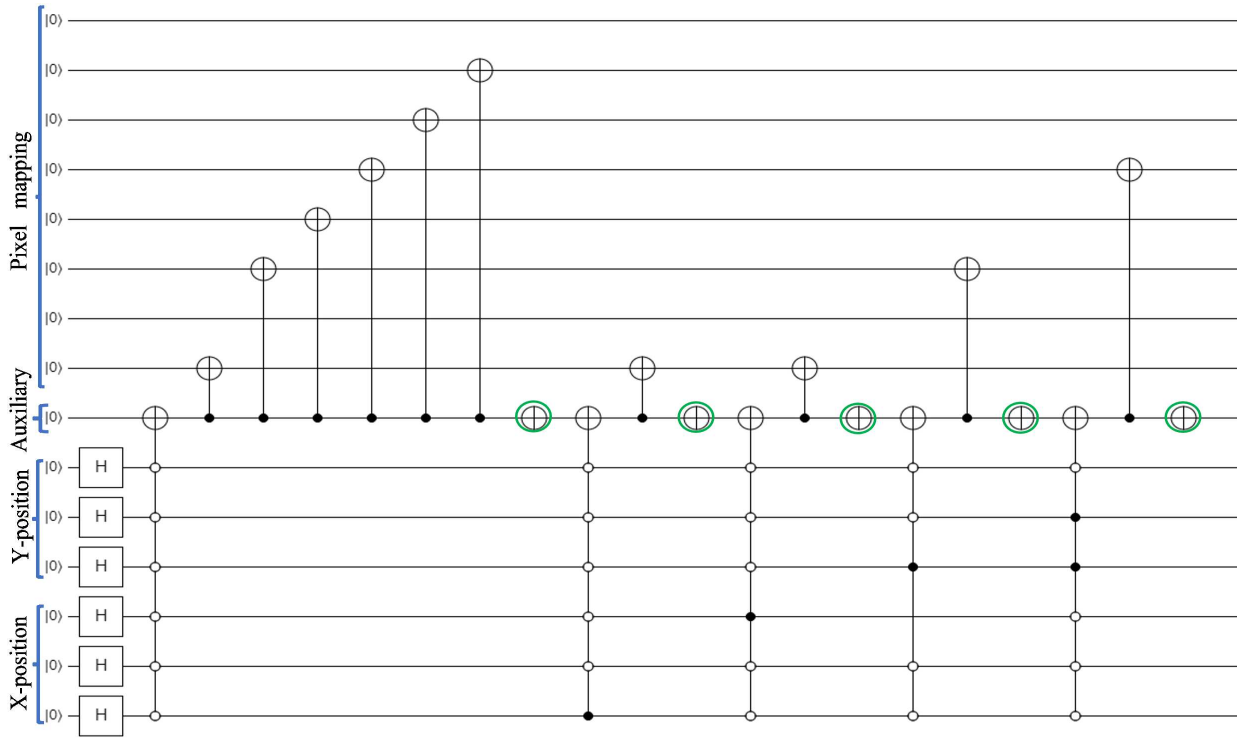}}
\caption{A SCMFRQI circuit diagram for pixel values representation}
\label{fig_proposed_Chematic_diagram}
\end{figure}

There has been a lot of work done for classical image and video compression and few are found in \cite{paul2005real,paul2018efficient}. In the quantum domain, an FRQI(flexible representation of the quantum image) was developed in \cite{b13} inspired by the pixel-wise representation of the classical image. It encodes the color and position of the image using the associated angle and kets of one qubit. It is only capable to represent four-pixel values of images. Entanglement-based image representation was proposed in \cite{b14}.\\ 
Jiang et al. proposed a compression method using GQIR(general quantum image representation) approach via the DCT approach based on JPEG (joint photography expert group) image \cite{pm}. A quantum-based equivalence pixel image from a bit pixel image was proposed in \cite{b15}. To address the classical color image representation issue, a NEQR was proposed in \cite{bb10}. Its resolves the FRQI approach limitations because it provides a way to represent the pixel-based gray-scale representation. For real-size images, it is still not clear how and in what way to represent them without tiny square images. To resolve the rectangular shape issue, INEQR (improved NEQR) approach was proposed in \cite{b9}.In the INEQR approach, how color and big-size images are represented is still not clear. A GQIR (general quantum image representation) approach was proposed in \cite{b18} which uses a logarithmic scale to represent the rectangular size of an image. The problem with the GQIR approach generates a lot of redundant bits. \\
An EFRQI (efficient flexible representations of quantum image) was proposed to decrease the state preparation complexity of GQIR and NEQR approaches \cite{b8}. Using the same Toffoli gate twice also generates a higher amount of required bits.  
Fig. \ref{fig_proposed_Chematic_diagram} shows the circuit diagram of the SCMFRQI scheme for representing 125(X=0,Y=0), 1(X=1,Y=0), 1(X=4,Y=0), 4(X=0,Y=1), and 16(Y=3,X=0) pixels values \cite{haque2022novel}. Rather than using the same Toffoli gate twice, SCMFRQI uses a single reset gate for each pixel or coefficient (shown in the green circle) \cite{bb17}. 

\section{Proposed Approach}
\label{P_M}
\begin{figure}[htbp]
\centerline{\includegraphics[width=\linewidth,height=10cm]{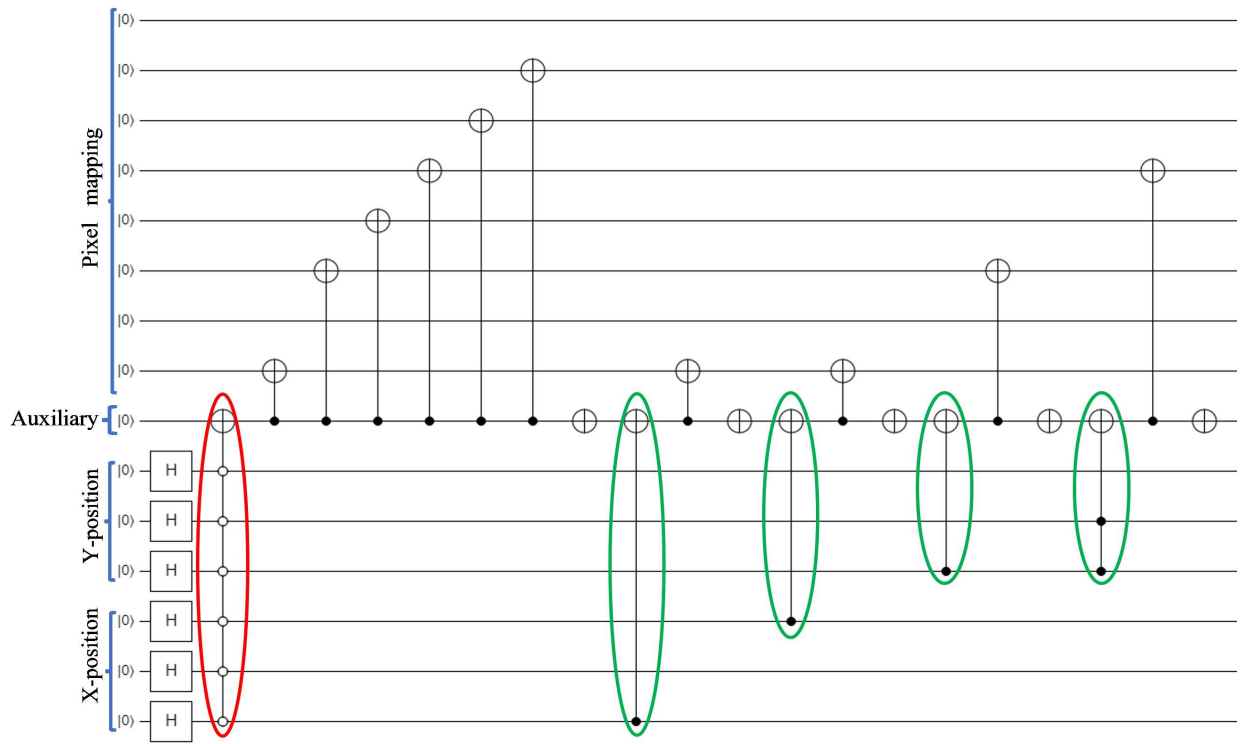}}
\caption{The proposed ZSCNEQR circuit diagram for pixel or coefficient values representation where an initial connection  (marked as red) and zero-discarded zone (marked as green) to save the state preparation bits}
\label{fig_proposed_SCMNEQR_diagram}
\end{figure}

Fig. \ref{fig_proposed_SCMNEQR_diagram} shows the circuit diagram of the proposed ZSCNEQR scheme for representing pixel values or coefficient values. Rather than using both '0' and '1' for state preparation, it uses '1' only for each pixel or coefficient connection (shown in the green circle). To map the quantum circuit, the Quirk simulation tool is used \cite{bb17}. 

A $2^a \times 2^b$ image size is considered for representation and compression purposes.\\ 
In the proposed ZSCNEQR approach, the steps involved are \\
Step 1: DCT and quantized. \\
Step 2: Prepare the quantized DCT coefficient. The proposed ZSCNEQR approach requires $q+2n+1$ qubits. Where $q$ is the number of required qubits to represent the pixel or coefficient values and $n=log_2(S)$. Where $S$ is the total number of blocks consisting of XY-positional blocks of the images. The initial state can be explained using the below equation \cite{b8}. 
 \begin{equation}
   |\Psi_0\rangle={\vert0\rangle}^{\otimes(q+2n+1)}
 \end{equation}
Then, $(q+1)$ identity gates and $2n$ Hadamard gates are used. In this step, the whole quantum step can be expressed as follows 
\begin{equation}
U=I^{\otimes{q+1}}\otimes H^{\otimes{2n}}
\end{equation}
$U$ transform $\Psi_0$ from initial state to intermediate state $\psi_1$.
\begin{equation}
\Psi_1=U(|\Psi_0\rangle)=I|0\rangle^{\otimes{q+1}}\otimes H^{\otimes{2n}}
\end{equation}
The final preparation step is done using $U_2$ quantum operator 
\begin{flalign}
\Psi_2&=U_2(|\Psi_1\rangle) \\  
&=\frac{1}{2^n} \sum_{i=1}^{2n-1}\sum_{j=1}^{2n-1}\,\left(|C_{YX}\rangle (|Y_{o}X_{o}\rangle \right)
+\frac{1}{2^n}|C_{Y_zX_z}\rangle|Y_{z}X_{z}\rangle) \nonumber
\end{flalign}
Where $|C_{YX}\rangle$ is the corresponding pixel value of the $Y_{o}X_{o}$ position which considers the values of one's only. On the other hand, $|C_{Y_zX_z}\rangle$  represents the zero position of pixels or coefficient, and $Y_{z}X_{z}$ represents its positional value. The quantum transform operator is $U_2$ is given below  
\begin{equation}
U_2=\prod_{X=0,....,2^n-1}\prod_{Y=0,....,2^n-1}\, \left(U_{Y_{z}X_{z}}+U_{Y_{o}X_{o}}\right)
\end{equation}
The quantum sub-operator $U_{YX}$ is also given below 
\begin{flalign}
U_{YX}&= \left(I\otimes \sum_{ij\neq YX} {|ji\rangle {\langle ji|}} \right) +\sigma_{YX} \otimes|YX\rangle {\langle YX|}
\end{flalign}
where $ |YX\rangle {\langle YX|}=|Y_{z}X_{z}\rangle {\langle Y_{0}X_{0}|}+|Y_{0}X_{0}\rangle {\langle Y_{0}X_{0}|}$. 
The $\sigma_{YX}$ is given below
\begin{equation}
    \sigma_{YX} =\otimes^{q-1}_{i=0}{\sigma^i_{YX}}
\end{equation}
The function of $\sigma^i_{YX}$ is setting the value of $i_{th}$ qubit of (YX)'s quantized DCT coefficient. \\
 Step 3: Store the quantized coefficient after performing $8X8$ block DCT.\\
 Step 4: Inverse quantization. \\
 Step 5: Inverse DCT. \\
 Step 6: Compute RDC. \\
 The required number of bits of Toffoli gate ($B_{T}$) connection is   
 \begin{equation}
          B_{T}= (log_2(S_X)+log_2(S_Y)+C_T)\otimes{N_{tcn}}
 \end{equation}
 The required number of reset gates ($B_{rg}$) is given as
 \begin{equation}
          B_{rg}= R_N\otimes{N_{tcn}}
 \end{equation}
 Where $R_N$ is the required number connection.  
 The total number of required bits while zero ($B_{z}$) is discarded as follows  
 \begin{equation}
          B_{s_0}= B_{T}+B_{rg}-B_{z}
 \end{equation}
 In the proposed circuit, zero (o volts) and one (5 volts) are considered binary bits. Discarding zero from the state connection, making the proposed circuit more efficient because for creating each zero connection one operation gate is required. To complete the whole image circuit, a  huge amount of operational gates is required if '0' is considered. That way, in the proposed method, discarding '0' saves a huge amount of operational gates and does not have any effect since it is implemented by a quantum identity gate, which can be ignored. 
 The total number of required bits is calculated as follows    
 \begin{equation}  {BR|_{MB}}=\left(q_{o}+S_{bit}+B_{s_0}+A_{bit}+B_{BPE}\right)/(1000*1000)
 \end{equation}
\begin{equation}
     {BR|_{BPE}}=N_{rb}*N_{cb}*(B_{rbc}+B_{rbr})/(1000*1000)
 \end{equation}
Where $q_{o}$ is the total number of ones that appeared in the transfer coefficient or pixel values. A $S_{bit}$ is the sign bit that represents the sign of the non-zero pixel or coefficient values. A $N_{tcn}$ is the total number of non-zero coefficient or pixel elements. A $S_X$ and $S_Y$ represent the X and Y-position of nonzero pixels or coefficients. An $A_{bit}$ is the required number of bits that come from the auxiliary qubits. And $BR|_{BPE}$ is the block position error that is introduced to locate the blocks outside of the first block position. $(B_{rbc}$ and $B_{rbr})$ are the required number of bits to locate missing row ($N_{rb}$) and column block($N_{cb}$). 

\section{Result and discussion}\label{R_D}

In this section, the experimental results are analyzed for deer$(1024\times1024)$, baboon $(512\times512)$, scenery $(512\times512)$ and peppers $(512\times512)$ images \cite{bb17,bb18} for performance measure of the proposed ZSCNEQR approach. To measure its performance, two types of experiments have been conducted: one is the direct approach and another one is the indirect approach via the DCT preparation approach.  
\subsection{Pixel domain representation analysis}
\begin{figure}
\centering
\includegraphics[width=0.5\textwidth, height=6cm]{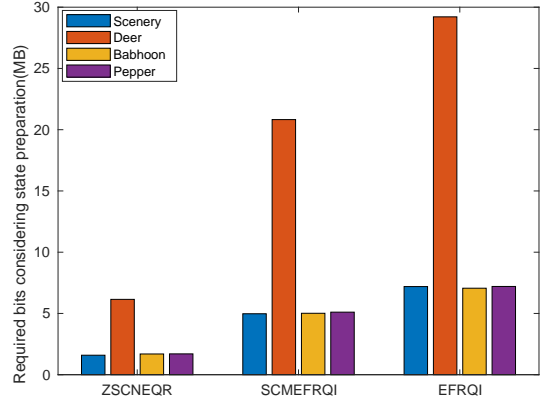}
\caption{Required bits for the proposed ZSCNEQR approach}
\label{B_PSNR_Four_image}
\end{figure}

Fig. \ref{B_PSNR_Four_image} shows the required number of bits for the grayscale of the considered deer, baboon, scenery, and pepper images of the proposed ZSCNEQR approach. \\
For scenery image,  Fig. \ref{B_PSNR_Four_image} shows the proposed ZSCNEQR approach requires 1.59 MB (megabits) where SCMEFRQI and EFRQI require 4.97 MB and 7.06 MB respectively. Therefore, the compression ratio of the scenery image in the case of the proposed ZSCNEQR approach compared to SCMEFRQI and EFRQI approaches are 3.12:1, and 4.44:1 respectively. \\
In the case of the deer image, Fig. \ref{B_PSNR_Four_image} depicts that the required number of bits of the proposed ZSCNEQR, SCMEFRQI, and EFRQI approaches are 6.15 MB,20.82 MB, and 29.21 MB respectively. The compression ratios of the proposed ZSCNEQR approach are 3.38: 1, and 4.74:1 respectively compared to SCMEFQI and EFRQI approaches. \\
On the other hand, in the case of the baboon image, Fig.   \ref{B_PSNR_Four_image} exhibits the required number of bits for the ZSCNEQR, SCMEFRQI, and EFRQI approaches are  1.69 MB, 5.01 MB, and 7.19 MB respectively. The compression ratios of the proposed ZSCNEQR approach are 2.96:1, and 4.17:1 respectively in terms of SCMEFRQI and EFRQI approaches. In the case of the pepper image, the required number of bits for the proposed ZSCNEQR, SCMEFRQI, and EFRQI approaches are  1.69 MB, 5.10 MB, and 7.20 MB respectively. Therefore, the compression ratios of the proposed ZSCNEQR approach are 3.00: 1, and 4.23:1 referring to SCMEFRQI and EFRQI approaches respectively.    
\subsection{Transfer coefficient data domain representation and compression analysis}
\begin{figure}
\centering
    \subfigure [deer image]
    {
        \includegraphics[width=0.5\textwidth, height=5cm]{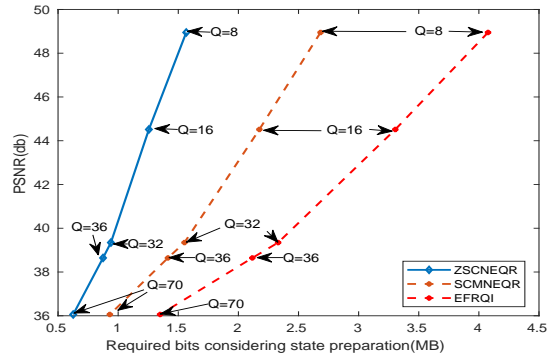}
        \label{red_channel_Deer}
    }
    \subfigure[baboon image]
    {
        \includegraphics[width=0.5\textwidth, height=5cm ]{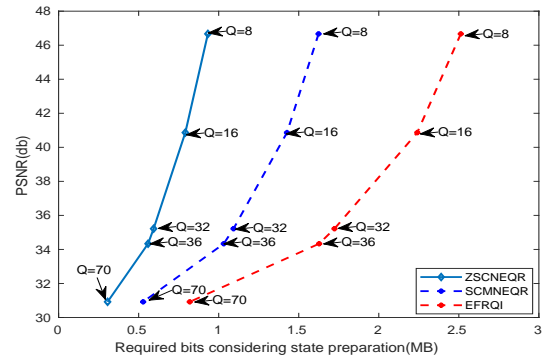}
        \label{red__channel_baboonsr}
    }
    \subfigure[scenery image]
    {
    \includegraphics[width=0.5\textwidth,height=5cm ]{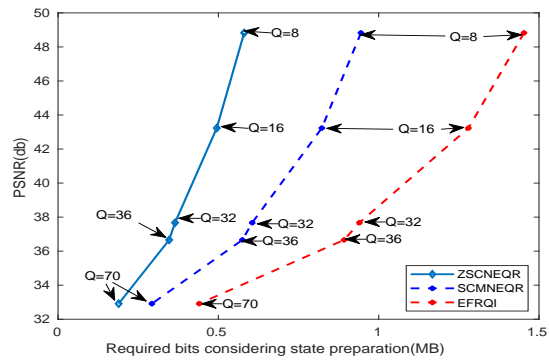}
        \label{red__channel_scenery}
    }
    \subfigure[peppers image]
    {
    \includegraphics[width=0.5\textwidth,height=5cm ]{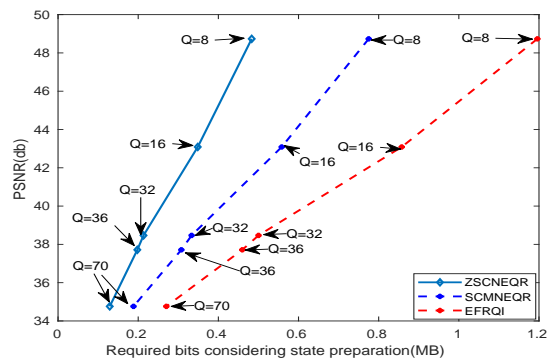}
        \label{red_channel_peppers}
    }
    \caption{Proposed scheme required bits versus PSNR for gray channel of considered images}
    \label{red_PSNR_comparison_color_image}
\end{figure}

Fig. \ref{red_PSNR_comparison_color_image} shows the comparison result of the proposed ZSCNEQR approach for the gray channel of the considered images. The comparison results in \ref{red_channel_Deer} show that the proposed ZSCNEQR approach draws the better rate-distortion curve(RDC) than both SCMEFRQI and EFRQI approaches respectively using Q=8, 16, 32, 64, and 70 quantization factors performed after DCT operation. For every quantization factor, the proposed ZSCNEQR approach provides lower bits compared to both SCMEFRQI and EFRQI approaches. 
Fig. \ref{red__channel_baboonsr} exhibits the computational result of RDC for baboon image in terms of proposed ZSCNEQR, SCEFRQI, and EFRQI approaches respectively. The comparison result shows that the proposed ZSCNEQR approach performed superior compared to SCMFRQI and EFRQI approaches. \\
On the other hand, Fig. \ref{red__channel_scenery} depicts the comparison result of the proposed ZSCNEQR approach which is better compared to SCMEFRQI and EFRQI approaches. In the meantime, SCMEFRQI, and EFRQI draw lower RDC values means both approaches required higher bits but provide the same PSNR values compared to the proposed ZSCNEQR approach. Fig. \ref{red_channel_peppers} exhibits the comparison result of RDC values of the proposed ZSCNEQR approach of the peppers image. The comparison result shows that for all quantization factors, the proposed ZSCNEQR approach performs more efficiently to represent and compress the gray channel of the peppers image. The result has shown that the proposed scheme is able to represent and compress the gray channel of the image efficiently compared to others.   



%
%
%
\section{Conclusion}\label{CC}
In this work, through the analysis of the literate survey, the limitations of the existing quantum models have been figured out. Then a novel, SCMEFRQI approach has been proposed for any size of image representation and compression to improve the EFRQI approach which is the latest and best existing model for state connection. it uses an auxiliary qubit and resets gate that modifies the state connection and stores and compresses the color image information of all the pixels or transfer coefficient values instead of probability. Another advantages is that, it also avoid the complex rotation operation. Besides, it is more capable to represent and compress the image compared with existing models.      
\section*{Acknowledgment}
The author declare that there is no conflict of interest.

\bibliographystyle{unsrt}  


\end{document}